\documentclass[twocolumn,prl,superscriptaddress,showpacs]{revtex4}

\usepackage{graphicx}
\usepackage{amsmath,amssymb}

\begin{document}

\title{Two-dimensional quantum liquids from interacting non-Abelian anyons}

\author{Andreas W.W. Ludwig}
\affiliation{Physics Department, University of California, Santa Barbara, California 93106}
\author{Didier Poilblanc}
\affiliation{Laboratoire de Physique Th\'eorique, CNRS and Universit\'e 
de Toulouse, F-31062 Toulouse, France}
\author{Simon Trebst}
\affiliation{Microsoft Research, Station Q,
University of California, Santa Barbara, CA 93106} 
\author{Matthias Troyer}
\affiliation{Theoretische Physik, ETH Zurich, 8093 Zurich, Switzerland}

\date{\today}

\begin{abstract}
A set of localized, non-Abelian anyons -- such as vortices in a $p_x + i p_y$ superconductor
or quasiholes in certain quantum Hall states -- gives rise to a macroscopic degeneracy.
Such a degeneracy is split in the presence of interactions between the anyons. Here we show that in two spatial dimensions this splitting selects a {\sl unique} collective state as ground state of the interacting many-body system. 
This collective state can be  a novel gapped quantum liquid nucleated 
inside the original parent liquid (of which the anyons are excitations). 
This physics is of relevance for any quantum Hall plateau realizing a non-Abelian quantum Hall state
when moving off the center of the plateau.
\end{abstract}


\maketitle

\noindent
While many conventional phases of matter appear as a consequence of spontaneous symmetry breaking, quantum systems may form more unusual ground 
states, so-called {\sl quantum liquids}, which do not break any symmetry \cite{Wen}. 
One particularly captivating case are two-dimensional {\sl topological} quantum liquids, which exhibit a bulk gap and harbor exotic quasiparticle excitations, so-called anyons, that obey fractional statistics \cite{LeinaasMyrheim}. 
Time-reversal symmetry breaking topological liquids with non-Abelian anyons have attracted considerable recent interest in the context of fractional quantum Hall states \cite{MooreRead},
rotating Bose-Einstein condensates \cite{Cooper01}, and
unconventional $p_x + i p_y$ superconductors \cite{ReadGreen}. 
When populating such a topological quantum liquid with a set of anyons, their non-Abelian nature manifests itself in a degenerate {\sl manifold of states} that grows exponentially with the number of anyons -- even in the case of fixed particle positions. 
It is this degenerate space of states which is key to proposals for topological quantum computation \cite{Nayak08}.

%
However, this degeneracy will be {\sl split by interactions} between the anyons. If the anyons are spaced far from each other, e.g. their distance is much larger than the coherence length, this interaction will be exponentially small and so will be the splitting.
In this manuscript, we will address the question of how this degeneracy will be lifted and a new collective ground state is formed when interactions cannot be ignored.
In particular, we determine the nature of this collective state for a two-dimensional arrangement of non-Abelian anyons. 
Note the conceptual similarity of this question with the one asked in the context of fractional quantum Hall states, where the Coulomb interaction splits the degenerate manifold of electronic states of a partially filled Landau level and leads to the formation of a new collective ground state -- the fractional quantum Hall state \cite{Laughlin83} .

\begin{figure}[b]
\begin{center}
  \includegraphics[width=\columnwidth]{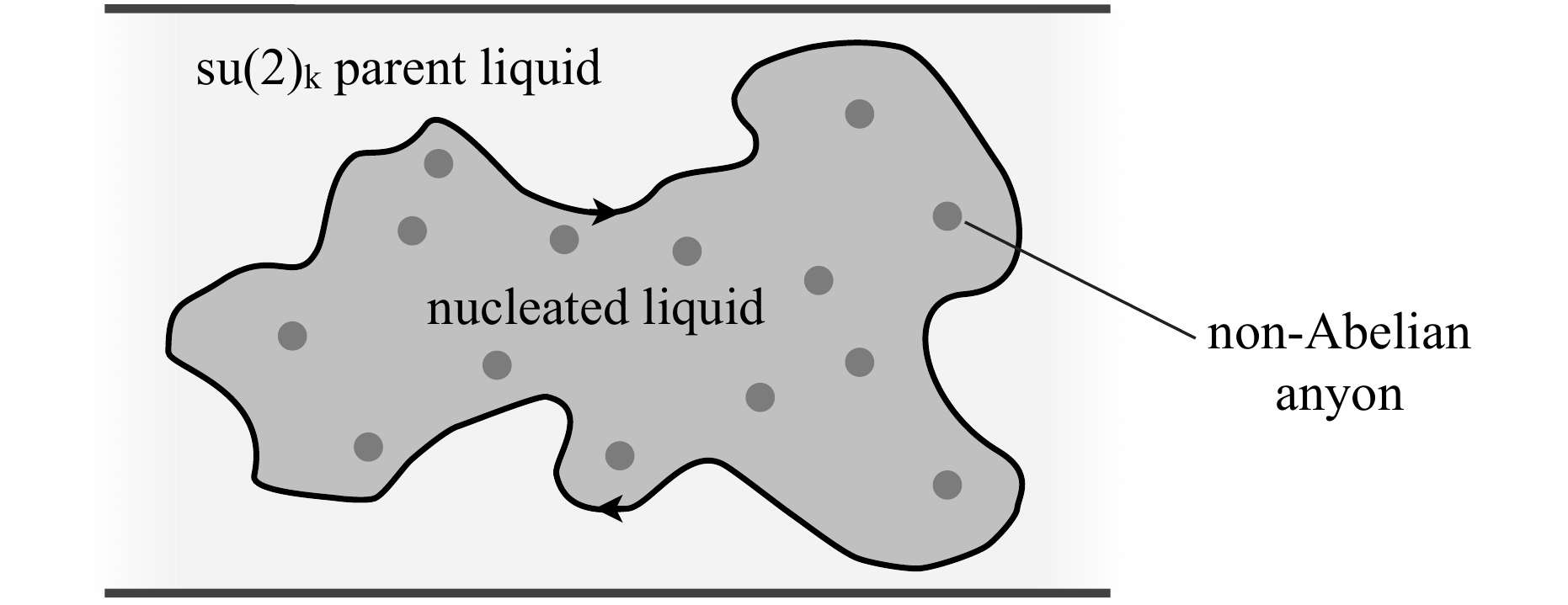}
\end{center}
\caption{
   The collective state of a set of interacting, localized, non-Abelian anyons is a gapped quantum liquid
   which is nucleated within the original parent liquid.
   The two liquids are separated by a neutral, chiral edge state.
 }
\label{Fig:NucleatedLiquids}
\end{figure}

%
One example of where this physics is of relevance are non-Abelian quantum Hall states {\sl off the center} of the quantum Hall plateau. The quantum Hall state at filling fraction $\nu=5/2$ is considered the best candidate to-date, with an intense experimental effort \cite{Dolev08, Radu08, Willet09} aimed at demonstrating its non-Abelian nature \cite{MooreRead}. 
By tuning the magnetic field away from the center of the plateau a finite density of quasiholes or quasiparticles will populate the quantum Hall liquid depending on whether the magnetic field is increased or decreased.
It is natural that these quasiholes (or quasiparticles) would form a Wigner crystal, which renders
them immobile. 
Remarkably, as we show in this manuscript, interactions between these anyons can then lead to the formation of a novel {\sl gapped} quantum Hall liquid
\footnote{Note that this quantum liquid forms ``on top" of the Wigner crystal of localized anyons, {\sl i.e.} the liquid forms only in the topological degrees of freedom without affecting the underlying crystalline positional order.},
which is separated from the original parent liquid by a neutral, chiral edge state as illustrated in
Fig.~\ref{Fig:NucleatedLiquids}.
These gapless edge modes contribute to heat transport leading to a distinct experimental signature.

To be specific, we consider here a setting where $su(2)_k$ anyons are pinned 
to static positions. 
These non-Abelian anyons are described by generalized angular momentum quantum numbers of so-called $su(2)_k$ Chern-Simons theories \cite{Witten}, which correspond to certain quantum deformations of SU(2) \cite{SU2q}. 
The case of $k=2$ describes the non-Abelian nature of anyonic excitations in the Moore-Read state,
as well as the Majorana modes of half-vortices in $p_x + i p_y$ superconductors.

The most elementary example of the lifting of a degeneracy occurs for a pair of non-Abelian anyons. 
An interaction mediated by topological charge tunneling \cite{Feiguin07,Bonderson09}
results in a splitting of the energies of the possible pair states
-- similar to the singlet-triplet splitting of two ordinary spin-1/2's.  
We consider the most relevant case where the anyonic degree of freedom corresponds to a generalized angular momentum $j=1/2$ of these $su(2)_k$ theories.
They obey the fusion rule $1/2 \otimes 1/2 = 0 \oplus 1$ reminiscent of two ordinary spin-1/2's coupling into a singlet or triplet. 
We therefore denote a coupling favoring the generalized $j=0$ (singlet) state as antiferromagnetic (AFM) coupling, while ferromagnetic (FM) coupling then corresponds to favoring the generalized $j=1$ (triplet) state. 
The magnitude and sign of this splitting are not universal and in general depend\ on microscopic details such as the spatial separation between the anyons and the nature of the topological liquid. 
Explicit calculations of this pair splitting have recently been performed for 
two quasiholes in a Moore-Read quantum Hall state \cite{Baraban09},
a pair of vortices in a $p_x + i p_y$ superconductor \cite{Cheng09},
and a pair of vortex excitations in the gapped ``B-phase" of Kitaev's honeycomb model 
in a magnetic field \cite{Lathinen08}.

\begin{figure}[t]
\includegraphics[width=\columnwidth]{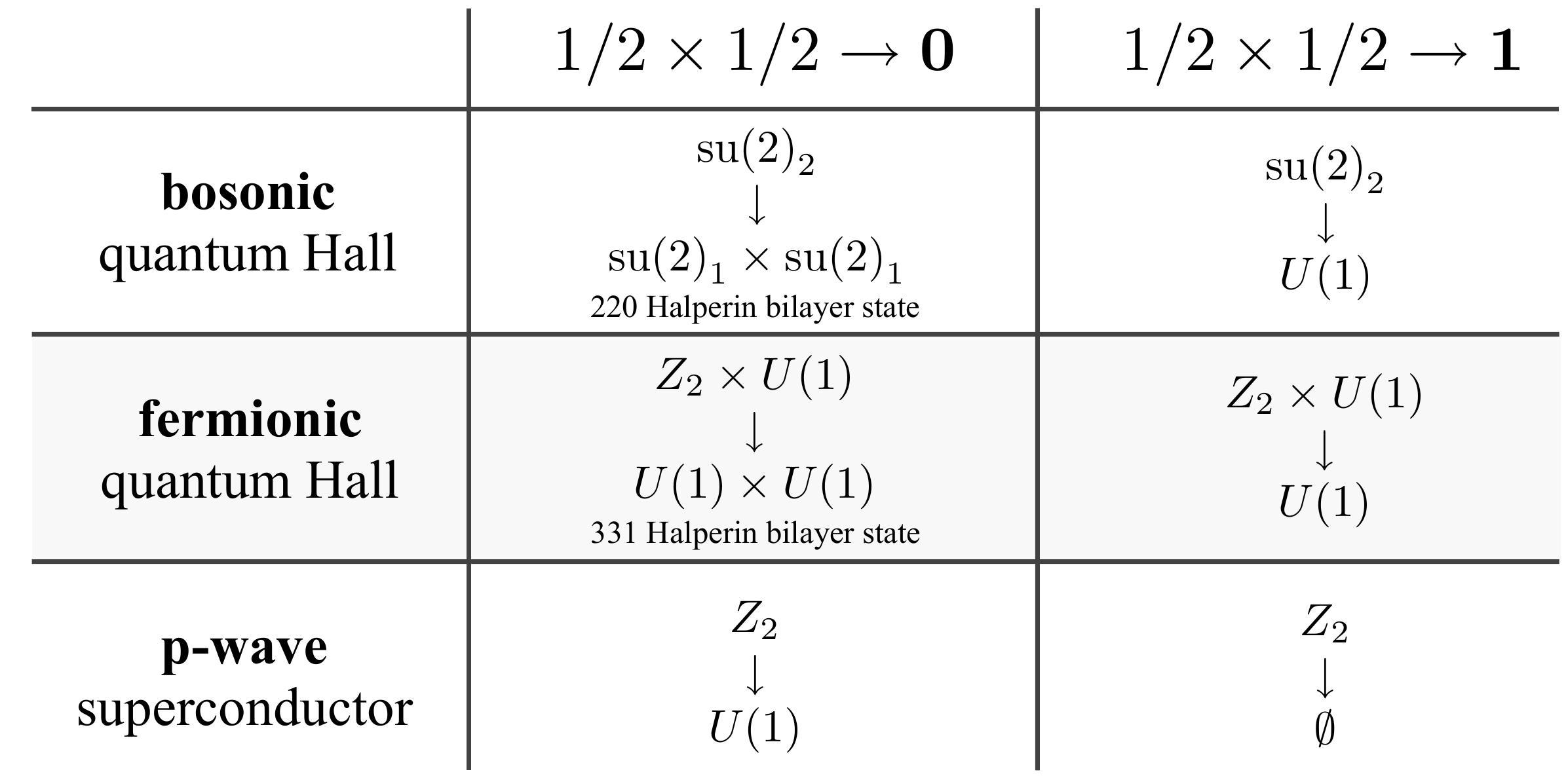}
\caption{
   Table of collective ground-states of interacting Majorana zero modes ($k=2$).
} 
\label{Fig:TableNucleatedLiquids} 
\end{figure}

For more than two anyons this elementary pair interaction will result in a splitting of their manifold
of macroscopically degenerate states and the formation of non-trivial collective quantum many-body states.
The main result of this paper is that the ground state of a two-dimensional array of interacting 
su(2)$_k$ anyons is a gapped non-Abelian su(2)$_{k-1}\times$su(2)$_1$ liquid for antiferromagnetic couplings and a gapped Abelian U(1) liquid for ferromagnetic couplings.
In Fig.~\ref{Fig:TableNucleatedLiquids} we discuss our results for the case of interacting
Majorana zero modes (corresponding to $k=2$) and apply them in the context of rotating Bose-Einstein condensates, the Moore-Read quantum Hall state and the $p_x+ip_y$ superconductors.
We note that some of these special cases have been analyzed before and are indeed captured
by our more general result, thereby further corroborating it.
The first column gives results for AFM anyon interactions, whereas the second column gives results for FM interactions.
Let us first focus on the first row `{\it bosonic quantum Hall}', which describes the non-Abelian physics of certain proposed rotating Bose-Einstein condensates \cite{Cooper01}
and literally corresponds to the $k=2$ case of our result.
In this case we find that for AFM interactions an Abelian topological liquid, 
described by su(2)$_1 \times$su(2)$_1$ Chern-Simons theory, is nucleated. 
This state is identical to the (bosonic) ``220 Halperin bilayer state'' \cite{HalperinStates}.
For FM interactions, an Abelian $U(1)$ liquid is nucleated and all non-Abelian aspects of the 
original su(2)$_2$ state disappear.
The second row `{\it fermionic quantum Hall}' generalizes the above result to the fermionic
version \cite{ReadRezayi} of the su(2)$_k$ state, which can also be described by $Z_k \times U(1)$ non-Abelian Chern-Simons theory \cite{footnote-su2k}.
For $k=2$ this state occurs, e.g., in the Moore-Read quantum Hall state \cite{MooreRead}.
For AFM interactions between the anyons, the nucleated liquid is an Abelian $U(1) \times U(1)$
liquid, which corresponds precisely to the ``331 Halperin bilayer state''
\cite{HalperinStates}.  
For FM interactions the $Z_2$ degree of freedom simply disappears upon nucleating
an Abelian U(1) liquid. 
This was first discovered using Majorana fermions in the context of random FM anyon interactions
\cite{Read00}.
Both cases were recently also discussed using different techniques in Ref.~\cite{LevinHalperin08}.
The third row describes the case of $p_x+ip_y$ superconductors. 
The non-Abelian nature of vortices in such superconductors is captured by a non-Abelian Z$_2$ theory \cite{footnote-su2k}.
For AFM pairwise interactions between vortices an Abelian $U(1)$ liquid is nucleated.
This was observed for a triangular vortex lattice using free fermion calculations 
\cite{Grosfeld06,KitaevPachos} and the observation of a non-trivial
Chern-number \cite{KitaevPachos}.
For FM vortex pair interactions, the topological order of the $p_x+ip_y$ superconductor 
is entirely destroyed. 

\begin{figure}[t]
\includegraphics[width=\columnwidth]{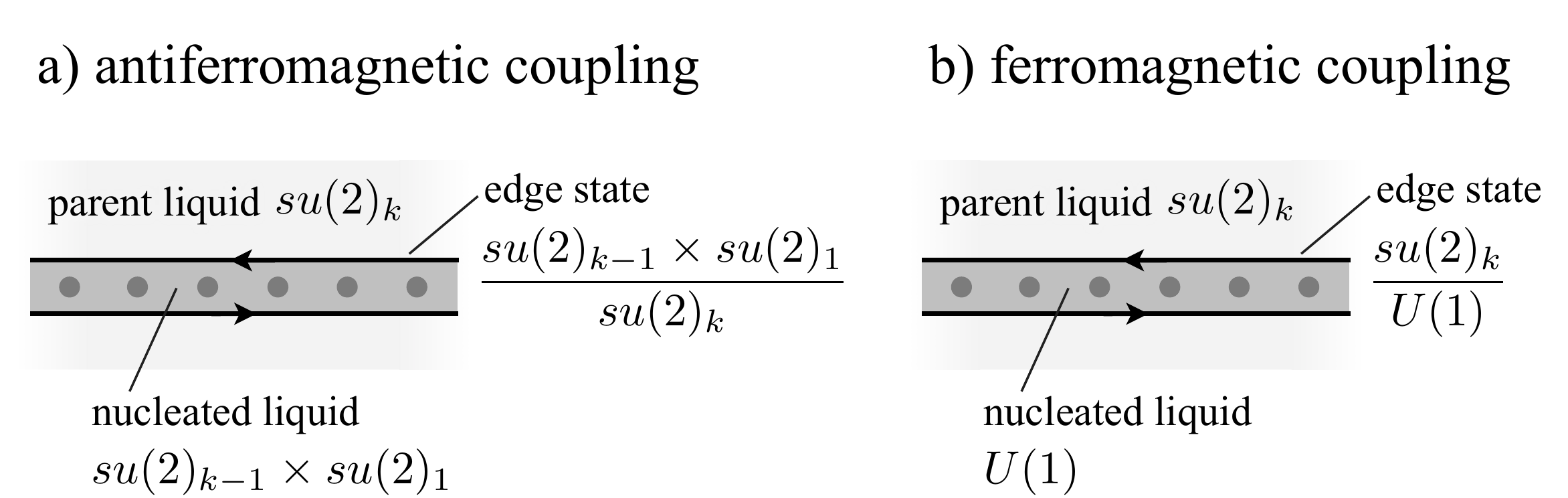}
\caption{
   A linear array of interacting su(2)$_k$ anyons nucleates a sliver of a
   gapped liquid separated from the original parent liquid by a chiral edge state.
} 
\label{Fig:1DLiquids} 
\end{figure}

The above case of $k=2$ turns out to be rather special, since this problem can be mapped to non-interacting Majorana fermions, which are more tractable than the general su(2)$_k$ theories for $k >2$.
A first step towards understanding the collective states of interacting su(2)$_k$ anyons has been taken by analyzing one-dimensional arrangements of localized anyons 
\cite{Feiguin07,Trebst08,Gils09}.
In this case, the gapless collective modes found for a chain of interacting anyons have been interpreted as
{\sl edge states}~\cite{Gils09} between the original topological liquid and a sliver of a novel, gapped nucleated liquid as illustrated in Fig.~\ref{Fig:1DLiquids}. 

\begin{figure}[t]
\includegraphics[width=\columnwidth]{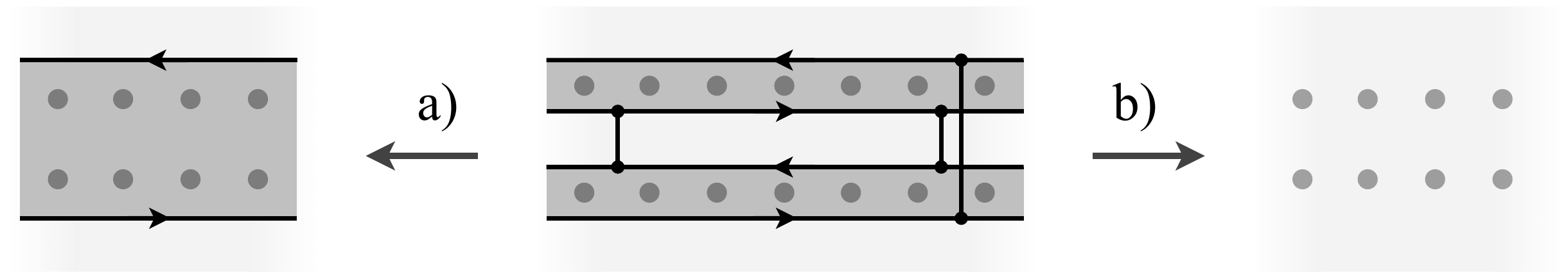}
\caption{
  Interactions between the anyons in two slivers can 
  a) gap out the inner edges and merge the two liquids into one droplet or
  b) gap out inner and outer edges thereby destroying the nucleated liquid in the sliver.
 }
\label{Fig:CoupledLiquids}
\end{figure}
\begin{figure}[t]
\includegraphics[width=\columnwidth]{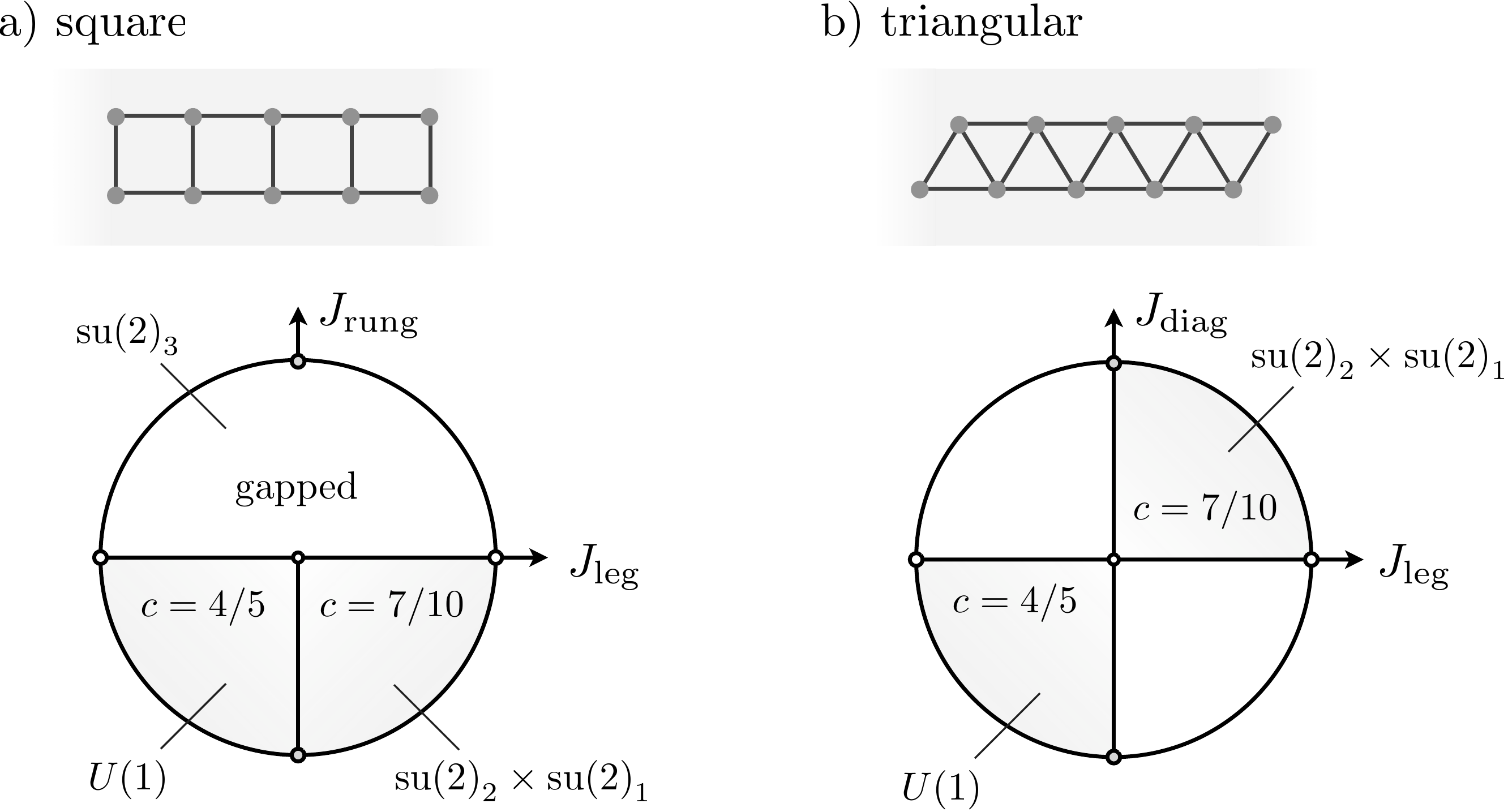}
\caption{
  The phase diagrams for anyonic su(2)$_3$ ladder systems with 
  a) square lattice geometry (of width $W=2$ or $W=4$) and
  b) triangular (of width $W=2$) 
  versus their respective couplings. 
  The shaded wedges indicate gapless phases, which are described by conformal field theories
  with central charges $c=4/5$ or $c=7/10$.
  These gapless theories correspond to edge modes between the original su(2)$_3$ anyonic liquid and a
  nucleated U(1) or su(2)$_{2}\times$su(2)$_1$ liquid, respectively.
 }
\label{Fig:PhaseDiagram}
\end{figure}

To approach the case of two-dimensional arrangements of interacting anyons we generalize
this ``liquids picture" and consider a system with several such one-dimensional slivers.
We find that this picture allows to precisely describe i) the subtleties in the spectrum of collective
edge states for a system of (multiple) decoupled slivers, and ii) the effect of coupling the individual slivers,
which originates from interactions between anyons in the different slivers. In particular, we find that
interactions between the slivers can gap out the {\sl inner} edges and 
merge the two slivers into one big droplet of nucleated liquid of identical character, 
see Fig.~\ref{Fig:CoupledLiquids}a). 
We numerically find that such droplet nucleation occurs for a wide range of couplings indicated by the shaded wedges in the phase diagrams of Fig.~\ref{Fig:PhaseDiagram} for different lattice geometries. Having established the merging of two slivers into a big droplet, this process can be iterated to yield macroscopically large droplets.

We first consider the case of two decoupled slivers, which gives rise to a sequence of five liquids and four intervening edges as illustrated at the center of Fig.~\ref{Fig:CoupledLiquids}.
It turns out that the combined spectrum
of gapless edge states is {\sl not} that of the free tensor product of the edge modes 
of the individual slivers, but rather is a subset of the latter.
This subset can be understood when considering all possible
tunneling processes between the various liquids, which give rise to certain constraints.
In particular, the spectrum of these four edges 
takes on the form 
$$
(\psi_L)^{i_1}_{j_1} \ (\psi_R)_{j_1}^{i_2} \ (\psi_L)^{i_2}_{j_2} \ (\psi_R)_{j_2}^{i_1} \,,
$$
where $(\psi_L)^{i}_{j}$ is a primary field in the coset \cite{gko86}
$su(2)_{k-1}\times su(2)_1/su(2)_k$ and describes the tunneling in and out of the
respective edge between the two topological liquids appearing in the coset
\footnote{The sequence $(i_1, j_1, i_2, j_2, i_1)$ is associated with the consecutive
five liquids in the center of Fig.~\ref{Fig:CoupledLiquids}.}.
Here $i$ is a quantum number in $su(2)_{k}$ and $j$ in $su(2)_{k-1}$.
Consider for simplicity odd $k$, in which
case we can choose $i$ to run over integer
values $0, 1, ..., (k-1)/2$ and $j$ run over values
$0,1/2, 1,..,(k-1)/2$.
No topological charges can be ejected into the surrounding su(2)$_k$ liquid in a physical
tunneling process, which enforces $i_1=0$.
With these constraints we can then derive the complete conformal spectrum of the four edges,
including conformal weights, degeneracies, and topological charge \cite{Feiguin07} assignments.
Checking these results against numerical calculations of the spectra we find perfect agreement.


We now turn to a microscopic lattice model of interacting anyons, which allows for a precise
numerical characterization of the collective ground states.
As a concrete realization we consider interacting $su(2)_3$ (Fibonacci) anyons arranged 
in a square lattice geometry, which we build up by coupling chains of anyons
into $W$-leg ladders.
We denote the strength of the anyon-anyon interaction as 
$J_{\rm leg}$ and $J_{\rm rung}$ for the coupling along and perpendicular to the chains, respectively, with positive (negative) coupling referring to AFM (FM) interactions favoring the generalized $j=0$ ($j=1$) state.
We parametrize these couplings as $\tan \theta = J_{\rm rung} / J_{\rm leg}$.


Following, at first, a similar route as in the discussion of SU(2) quantum spin ladders \cite{OrdinaryLadders} we consider the anyonic ladders in their strong rung coupling limit $|J_{\rm rung}| \gg |J_{\rm leg}|$.
For an isolated rung the total spin of the ground state is 0 or 1/2 
\footnote{In the su(2)$_3$ theory spin 1 can be identified with $1/2$ and spin $3/2$ with $0$.}
depending on the sign of the coupling and the width of the ladder: 
it is a spin $1/2$ for odd width ladders with $J_{\rm rung}>0$, and for ladders with $J_{\rm rung}<0$ whose width is not a multiple of 3.
The low-energy effective model for weakly coupled rungs is that of a single spin-1/2 anyonic chain.
Indeed we find that the low-energy spectrum is gapless and can be described  by a conformal field
theory identical to the one of a single chain \cite{Feiguin07}.
In the particular case of $k=3$ 
the gapless theories are those of the tricritical Ising model ($c=7/10$), see also Ref.~\cite{Kareljan08}, or 3-state Potts model ($c=4/5$) for AFM and FM couplings along the chains, respectively. 
We numerically find that these gapless phases extend all the way to the weak rung coupling limit $|J_{\rm rung}| \ll |J_{\rm leg}|$ as shown in the phase diagram of Fig.~\ref{Fig:PhaseDiagram}a). 
Interpreting the gapless modes in these phases as edge states, we conclude that AFM couplings
along the chain result in the nucleation of an su(2)$_2\times$su(2)$_1$ topological liquid, while
for FM couplings 
a U(1) topological liquid is nucleated.
We note that for a triangular arrangement of anyons, 
characteristic for the positional ordering in a Wigner crystal, the nucleation of new liquids occurs for {\sl equal} signs of the couplings, see  Fig.~\ref{Fig:PhaseDiagram}b). This is in line with results obtained for $k=2$ on an infinite triangular lattice \cite{Grosfeld06}.

If, on the other hand, the total spin of the ground state of an isolated rung is 0,
the elementary excitation on a rung is gapped and carries  spin 1. Its gap remains finite when turning on a leg coupling $J_{\rm leg}$ and we find numerically that, as in ordinary SU(2) ladders, the gapped phase extends all the way to the weak rung coupling limit 
$|J_{\rm rung}| \ll |J_{\rm leg}|$.
The absence of gapless modes in this coupling regime indicates that the original su(2)$_k$ liquid
survives.
We summarize the phase diagram for ladders of width $W=2$ and $W=4$ in Fig.~\ref{Fig:PhaseDiagram}a).


\begin{figure}
\includegraphics[width=\columnwidth]{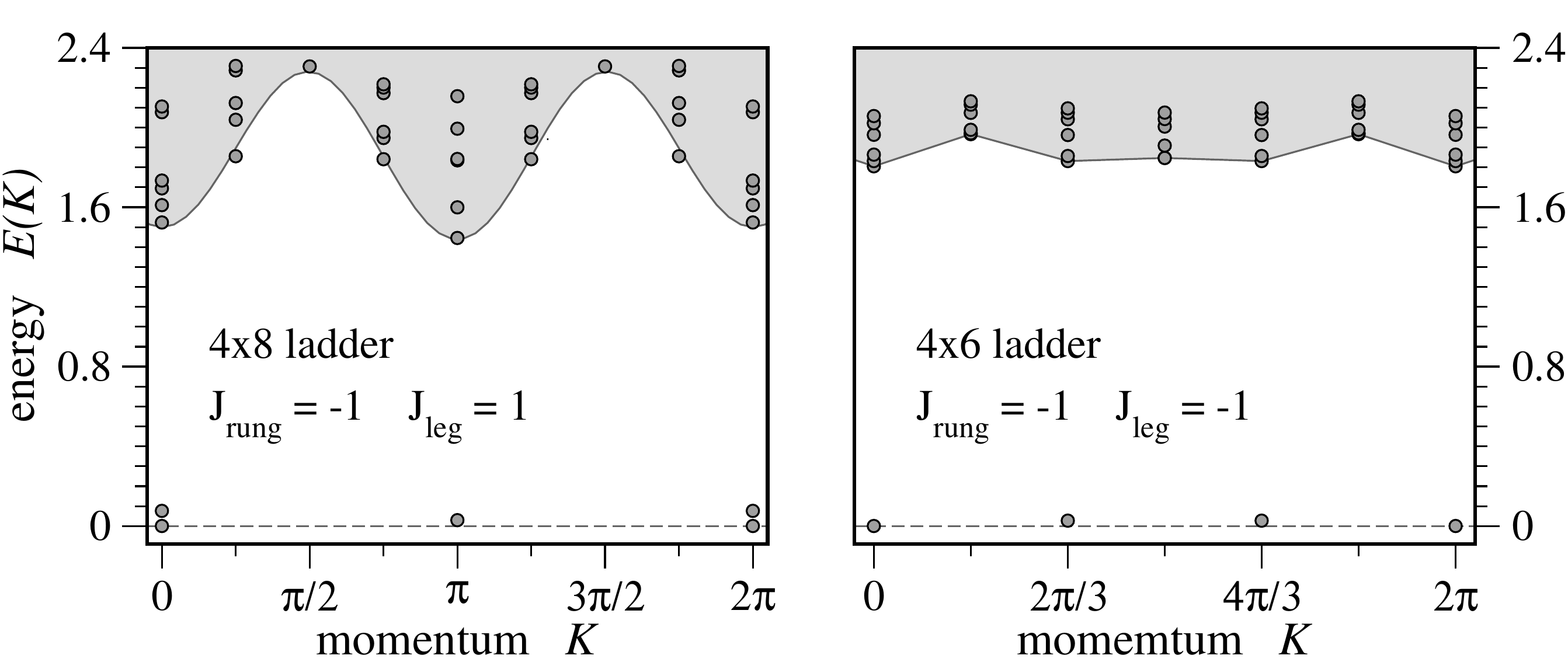}
\caption{
  Gapped energy spectra of 4-leg su(2)$_3$ anyonic ladders in a toroidal geometry
  versus the momentum along the legs.
 }
\label{Fig:BoundaryConditions}
\end{figure}

A characteristic feature of a topological quantum liquid is that it is sensitive to the topology of the underlying manifold \cite{Wen89}, which is reflected in the ground-state degeneracy and the occurrence of gapless edge modes for open boundaries. 
So far we considered ladder systems with open boundary conditions along the rung direction (an annulus) and understood the occurring gapless modes as edge states. Gluing the boundaries of this annulus to form a torus by adding a coupling between the first and last leg removes the boundary and is thus expected to also remove the edge states.
Indeed the spectrum of the 4-leg ladder on a torus is found to be gapped, 
as shown in Fig.~\ref{Fig:BoundaryConditions}. 
Further, the topological nature of the collective ground state is confirmed by the numerically observed degeneracy.
For both signs of leg coupling this degeneracy is found to be three-fold, which precisely corresponds to the expected number of different topological charges on the torus for the su(2)$_{2}\times$su(2)$_1$ theory
arising for $J_{\rm leg}>0$, and the U(1) theory arising for $J_{\rm leg}<0$.

\paragraph{Outlook.--}
Our essential result is that anyon-anyon interactions split the degeneracy of a macroscopic number of non-Abelian anyons, pinned at fixed spatial locations, and select a {\sl unique} gapped collective many-body ground state. 
For a wide range of anyon couplings, this
collective state is found to be again a topological
liquid, distinct from the original liquid of which the anyons are excitations.
This physics should be of relevance for any quantum Hall plateau realizing
a non-Abelian quantum Hall state: away from the center of the plateau, a
finite density of quasiholes or quasiparticles is expected to be present.
If these charged quasiparticles(-holes)  are pinned at fixed spatial
positions (e.g. by the formation of a Wigner crystal), 
our results predict the appearance of a {\sl different} quantum Hall state
with the {\sl same} electrical, but different thermal transport properties.
The latter arise solely from the splitting of the topological degeneracy due to
inter-anyon interactions. Furthermore, our results give a general many-body
perspective of how topological quantum computing schemes will fail at a
temperature scale comparable to the interactions between the anyons. While
the pairwise interaction between the anyons has been expected to be a
cause of decoherence for a topological quantum computer \cite{Bonderson09,
Cheng09}, our results demonstrate that at temperatures below the gap scale
of the nucleated liquid -- which we find to be of the same order as the
bare 2-anyon interaction -- all anyonic quasiparticles effectively
disappear by forming a new collective state thereby rendering
computational schemes based on the individual manipulation of anyons
impracticable.

\begin{acknowledgments}
A.W.W.L. was supported, in part, by NSF DMR-0706140, D.P.  by the French National Research Agency, and M.T. by the Swiss  National Science Foundation. We acknowledge support of the Aspen Center for Physics.

\end{acknowledgments}


\begin{thebibliography}{00}
\bibitem{footnote-su2k}
All non-Abelian aspects are incorporated in the $Z_k$ factor. This factor also appears when we represent \cite{Zamolodchikov} su(2)$_k$ as $Z_k \times U(1)$.
\end{thebibliography}
\end{document}